\documentclass[aps,prb,fleqn,twocolumn,showpacs, showkeys, superscriptaddress]{revtex4-1}% Physical Review B

\newcommand{\hofeb}{\mbox{HoFe$_{3}$(BO$_{3}$)$_4$}}
\newcommand{\rfeb}{\mbox{$R$Fe$_{3}$(BO$_{3}$)$_4$}}

\newcommand{\sisi}{\mbox{$\sigma$-$\sigma'$}}
\newcommand{\sipi}{\mbox{$\sigma$-$\pi'$}}
\newcommand{\pone}{\mbox{$P_{1}'$}}
\newcommand{\ptwo}{\mbox{$P_{2}'$}}
\newcommand{\tn}{\mbox{$T_N$}}
\newcommand{\tsr}{\mbox{$T_{SR}$}}
\newcommand{\lthree}{\mbox{$L_3$}}
\newcommand{\degree}{\mbox{$^{\circ}$}}

\usepackage{hyperref}

\usepackage{amssymb,amsmath} 
\usepackage{graphicx, amssymb}
\usepackage[dvips]{epsfig}
\usepackage{color}
\usepackage{epstopdf}

\begin{document}

\title{Ho and Fe magnetic ordering in multiferroic \hofeb\ }

\author{D. K. Shukla}
\email{dinesh.kumar.shukla@desy.de}
\affiliation{Deutsches Elektronen-Synchrotron DESY, 22607 Hamburg, Germany}
\author{S. Francoual}
\affiliation{Deutsches Elektronen-Synchrotron DESY, 22607 Hamburg, Germany}
\author{A. Skaugen}
\affiliation{Deutsches Elektronen-Synchrotron DESY, 22607 Hamburg, Germany}
\author{M. v. Zimmermann}
\affiliation{Deutsches Elektronen-Synchrotron DESY, 22607 Hamburg, Germany}
\author{H. C. Walker}
\affiliation{Deutsches Elektronen-Synchrotron DESY, 22607 Hamburg, Germany}
\author{L. N. Bezmaternykh}
\affiliation{L.V. Kirensky Institute of Physics, Siberian Branch of Russian Academy of Sciences, Krasnoyarsk 660036, Russia}
\author{I. A. Gudim}
\affiliation{L.V. Kirensky Institute of Physics, Siberian Branch of Russian Academy of Sciences, Krasnoyarsk 660036, Russia}
\author{V. L. Temerov}
\affiliation{L.V. Kirensky Institute of Physics, Siberian Branch of Russian Academy of Sciences, Krasnoyarsk 660036, Russia}
\author{J. Strempfer}
\affiliation{Deutsches Elektronen-Synchrotron DESY, 22607 Hamburg, Germany}

\date{\today}

\begin{abstract}

Resonant and non-resonant X-ray scattering studies on \hofeb\ reveal 
competing magnetic ordering of Ho and Fe moments. Temperature and 
X-ray polarization dependent measurements employed at the Ho \lthree\ edge directly reveal a 
spiral spin order of the induced Ho moments in the $ab$-plane propagating along the $c$-axis, a screw-type magnetic structure. 
At about 22.5 K the Fe spins are observed to rotate within the basal plane inducing spontaneous electric polarization, $\boldsymbol P$. Components of $\boldsymbol P$ in the basal plane and along the $c$-axis can be scaled with the separated  magnetic X-ray scattering intensities of the Fe and Ho magnetic sublattices, respectively. 

\end{abstract}

\maketitle
%77.22.Ej Polarization and depolarization
%75.30.Kz Magnetic phase boundaries (including magnetic transitions, metamagnetism etc.
%75.50.Ee Antiferromagnetics
%75.80.+q Magneto-mechanical and magneto-electric effects, magnetostriction
%78.70.Ck x-ray scattering
%61.05.cp x-ray diffraction
%64.70.-p Specific phase transitions
%75.25+z { Spin arrangements in magnetically ordered materials (including neutron and
%spin-polarized electron studies, synchrotron-source X-ray scattering, etc.).
\section{Introduction}
Spontaneous electric polarization in a magnetically ordered phase and a coupling between the two order parameters defining a magnetoelectric
behavior have been observed in many materials \cite{Che07}, and understood phenomenologically \cite{Mos06,Hos05}. However, recent observations of improper ferroelectricity \cite{Joh12,Lee11}, anomalous even within the realm of magnetoelectricity, require an even better understanding of the subject. Rare earth iron borates \rfeb, where $R$ = rare earth~\cite{Yen06,Zve06,Cha09,Kad10}, are multiferroics, characterized by long range magnetic wave vectors $q$$\|$$c$*~\cite{Ham10,Mo08,Rit08,Rit07,Fis06,Rit10}, and show a large magnetoelectric effect ($\boldsymbol P\sim100\mu Cm^{-2}$). The \rfeb\ crystal structure \cite{Kli05,Cam97} allows a dominant Fe-Fe exchange interaction, which is reflected by a \tn\ in a narrow temperature range (30-40 K) for compounds with different $R$. The rare earth exchange interaction takes place via O and B i.e. R-O-B-O-R. However, there is a more direct superexchange path between Ho and Fe through the R-O-Fe chains, causing $f-d$ exchange interaction. Evidence of the $f-d$ exchange interaction has been reported for several of these compounds, based on the observation of splitting of the ground-state doublet crystal-field level of Kramer's R$^{3+}$ ions using optical spectroscopy~\cite{Sta07}. A spin reorientation at low temperature is shown by the $R$= Gd \cite{Mo08} and Ho~\cite{Yen06,Cha09} compounds due to the large magnetic moment at the R site with a different magnetic anisotropy (usually easy-axis) than at the Fe site (easy-plane). However in the case of $R$ = Tb \cite{Rit07}, the rare earth anisotropy plays such a strong role that it already achieves an easy-axis magnetic structure at \tn. 

The key question regarding the spin arrangement in the basal plane of the aforementioned easy-plane ferroborates, which is essential for understanding the induced electric polarization, has remained unanswered. To understand the magnetic structure of these compounds neutron powder diffraction on $R$= Y, Pr, Nd, Tb, Ho and Er~\cite{Rit08,Rit07,Fis06,Rit10}, resonant magnetic X-ray scattering on $R$= Gd \cite{Mo08} and non-resonant magnetic X-ray scattering on $R$= Nd, Gd, Tb and Y \cite{Ham10} have
been performed. From neutron powder diffraction it has not been possible to determine the spin direction in the basal plane, since these compounds have a trigonal crystal structure. Concerning the X-ray scattering measurements on these compounds, until now, it has not been possible to directly determine the magnetic anisotropy of the rare earth, as the measured intensity variation as a function of the azimuthal angle about the scattering vector is flat~\cite{Mo08,Ham10}. Also resonant soft X-ray scattering measurements at the Fe L$_{2,3}$ absorption edges, which in principle could directly probe the Fe magnetic anisotropy, are not possible, as already the lowest indexed magnetic reflection (0 0 1.5) in these compounds is not accessible at these energies.

\hofeb\ undergoes a structural transition from R32 to P3$_1$21 ($T_S$ $\sim$ 427 K) and shows spontaneous polarization and magnetoelectricity below $\sim$~23~K, while the long range magnetic order sets in at \tn = 38 K \cite{Yen06,Hin03,Cha09,Pan09}. A spin re-orientation occurs at low temperature (\tsr $\sim$ 5 K) \cite{Yen06,Cha09}. Below \tsr, the polarization drops to zero \cite{Cha09}. We here combine resonant X-ray scattering measurements at the Ho \lthree-edge and high-energy (non-resonant) X-ray magnetic scattering (HEXMS) measurements at 100 keV, which allow the Ho and the Fe magnetic anisotropies to be directly determined. Full X-ray polarization analysis has been performed at the Ho \lthree\ resonance, from which an investigation of the Ho moments within  the basal plane becomes possible. HEXMS measurements are used to investigate the combined Ho and Fe moments oriented in the $ab$ plane. The Fe moments are found to rearrange in the $ab$ plane at around 22.5 K inducing magneto-electric behavior. The purpose of the present X-ray scattering study is to disentangle the Ho and Fe magnetic ordering and attempt to understand the associated spontaneous electric polarization.

\section{Experimental details}
Single crystals of \hofeb\ were grown using the flux method \cite{Bez05}. 
The resonant X-ray scattering (RXS) measurements at the Ho \lthree-edge were carried out in vertical scattering geometry at beamline P09 at the PETRA III storage ring at DESY Hamburg, Germany. The sample was mounted on a Psi-diffractometer so that $c\|\boldsymbol U_{3}$ (see Fig.~\ref{scattgeo}). For full polarization analysis, variable linearly polarized incident X-rays were generated using two 400 $\mu$m diamond quarter-wave plates in series~\cite{Sca09}. The polarization of the scattered signal was analyzed using a PG(006) analyzer crystal. Fig.~\ref{scattgeo} represents the scattering configuration at P09. 
The HEXMS investigations were carried out in horizontal scattering geometry, in transmission, at beamline BW5 at the DORIS III storage ring at DESY at a photon energy of 100 keV \cite{Bou98}. The absorption length in \hofeb\ is 2.12 mm at 100 keV, which allows one to probe directly the bulk of the sample in a similar fashion to neutron measurements. The beam size was 1~x~1~ mm$^{2}$. At high photon energies the magnetic scattering cross-section does not depend on the polarization of the X-rays \cite{Str96}.

\section{Results and discussion}
\begin{figure}[hbt]
\centering
\includegraphics[width=0.5\textwidth,trim=0.3cm 2cm 0cm 0.5cm, clip=true]{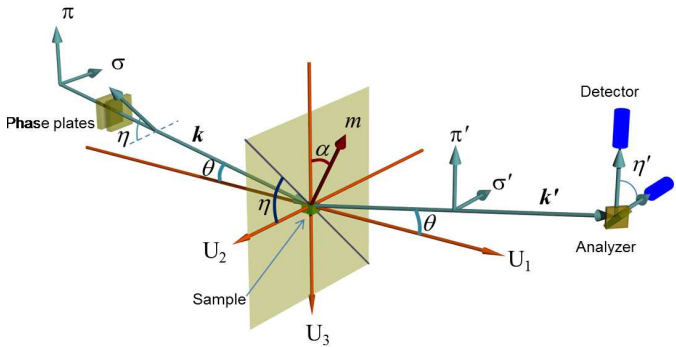}
\caption{(Color online) Vertical scattering geometry at beamline P09. $\theta$ is the Bragg angle, $\alpha$ is the angle between the magnetic moment $m$ and the scattering vector. $\eta$ and $\eta$$'$ define the angle of the incident polarization and the rotation of the analyzer crystal, respectively. $\boldsymbol k$ and $\boldsymbol k'$ are the incident and scattered wave vectors. $\sigma$($\sigma$$'$) and $\pi$($\pi$$'$) denote the polarization perpendicular and parallel to the diffraction plane of the incident (scattered) beam, respectively. $\boldsymbol U_{1}$, $\boldsymbol U_{2}$ and $\boldsymbol U_{3}$ define a basis for the magnetic structure \cite{Blu88}.}
\label{scattgeo}
\end{figure}

\begin{figure}[hbt]
\centering
\includegraphics[angle=0,width=0.5\textwidth,trim=0.3cm 0.1cm 0cm 1cm, clip=true]{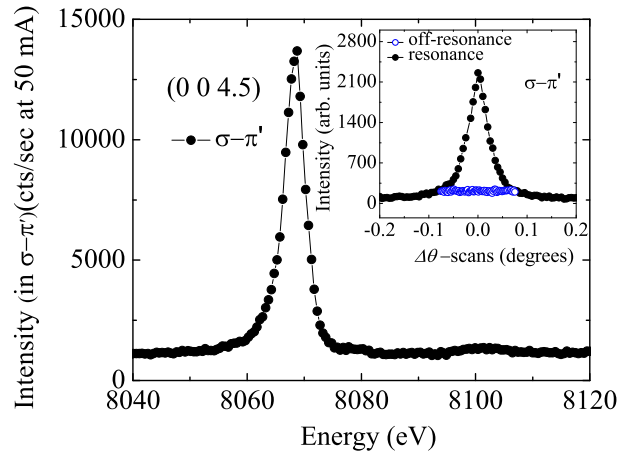}
\caption{(Color online) Energy dependence at the magnetic (0 0 4.5) reflection in the \sipi\ channel at T $\sim$ 6 K. The inset shows the rocking curves of the same reflection at resonance as well as off-resonance in the \sipi\ channel.}
\label{edep}
\end{figure}

In Fig.~\ref{edep}, the energy dependence of the intensity of the (0 0 4.5) reflection at 6~K in the \sipi\ channel is shown. The resonant enhancement at the Ho \lthree-edge in the \sipi\ channel confirms
the magnetic origin of this reflection.  Non-resonant (at energies away from the Ho \lthree-edge) intensity was observed in neither the \sisi\ nor the \sipi\ channel at this position, as seen from the inset of Fig. \ref{edep}. 

Full X-ray polarization analysis of the magnetic (0 0 4.5) reflection has been performed at the Ho \lthree\ resonance at 6~K. The incident linear polarization is varied around the incident wave vector $\boldsymbol k$ using phase plates and at the same time the polarization of the scattered beam is analyzed using a polarization analyzer. For each incident polarization state, rocking scans of the analyzer crystal at different analyzer positions $\eta'$ (with steps of 25\degree\ between 0\degree\ to 150\degree) are performed upon rotating the analyzer and detector assembly around the scattered wave vector $\boldsymbol k'$ (see Fig.~\ref{scattgeo}). The measured intensities are fitted by the following relation \cite{Maz07}:

\begin{equation}
  \label{ipol}
  I(\eta,\eta') = \frac{I_{0}}{2}(1+P_{1}'(\eta)cos2\eta'+P_{2}'(\eta)sin2\eta')
\end{equation}

\begin{figure}[hbt]
\centering
\includegraphics[angle=0,width=0.5\textwidth,trim=1.5cm 1.0cm 0cm 1.5cm, clip=true]{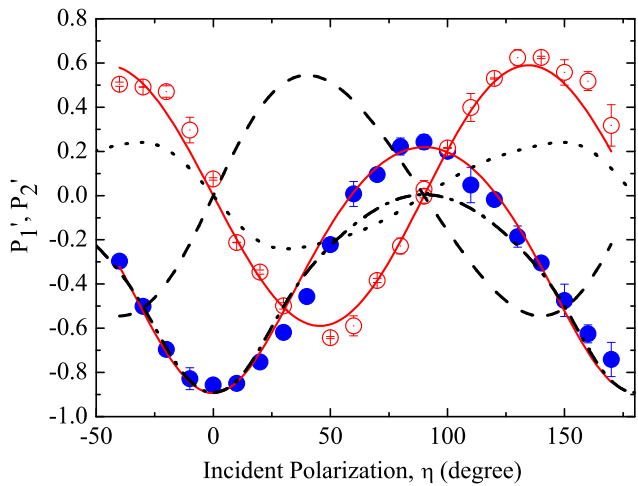}
\caption{(Color online) \pone\ (solid circles) and \ptwo\ (hollow circles) measured at the magnetic (0 0 4.5) refection at resonance as a function of the incident polarization at T = 6~K. Solid lines are fits corresponding to the Ho moments forming an $ab$-plane spiral along the $c$-axis. The dashed line is the simulated \ptwo\ variation when moments are assumed to be oriented along the $c$-axis. The dashed-dotted and dotted lines are simulated \pone\ and \ptwo\ variations, respectively, when there are equally populated magnetic domains with moments in the $ab$-planes. Our data therefore conclusively excludes these alternative hypotheses for the absence of any azimuthal variation in the intensity.}   
\label{poldep}
\end{figure}

$I_{0}$ and the Poincar$\acute{e}$-Stokes parameters, \pone\ and \ptwo\ , are the fitting parameters. \pone\ and \ptwo\ define the linear polarization state of the scattered X rays \cite{Sto52}. Figure~\ref{poldep} shows \pone\ (solid circles) and \ptwo\ (hollow circles) measured for the magnetic (0 0 4.5) reflection as a function of incident polarization angle $\eta$. 
It can be seen that \pone\ is a minimum for $\sigma$ incident light and a maximum for $\pi$ incident light, opposite to the behaviour of a charge peak. In order to fit the parameters \pone\ and \ptwo\ we use the density matrix formalism \cite{Blu88} and the matrix for the resonant magnetic cross-section as introduced by Hill and McMorrow \cite{Hil96}. Only electric dipole processes are considered.  

To fit the experimental data, instead of directly assuming  an $ab$ plane spin spiral, a formalism derived for a more general $c$-axis conical spiral structure was used \cite{Hil96}. The best fit of \pone\ and \ptwo\ for the magnetic (0 0 4.5) reflection is obtained when Ho moments are forming a basal plane spin spiral around the $c$ axis. In this crystal structure (s.g. P3$_1$21) the wave vector (0 0 3/2) representing a doubling of the unit cell along the $c$ axis allows the realization of a basal-plane spiral with a rotation of the Ho moment of 60 degrees from one crystallographic plane to another crystallographic plane. This is in agreement with the absence of an intensity variation with the variation of azimuth as observed in previous X-ray scattering studies \cite{Ham10,Mo08}, which is observed when either spins are aligned along the scattering vector ($\|$ to the $c$ axis) or form a spin spiral structure around it. For these two extreme cases a significant difference is expected for $\ptwo$ \cite{Blu88}. \ptwo\ should have a maximum at $\eta = 45\degree$ for spins oriented along the $c$ axis (dashed line in Fig.~\ref{poldep}). Sometimes naturally occurring defects i.e. magnetic domains might also cause an insensitivity of the resonant magnetic scattering to azimuthal rotation. To rule out such a possibility we have also simulated the full polarization analysis results for magnetic domains. For the simulation, moments are considered to have an ab-plane collinear magnetic structure i.e. there are three (six) equally populated domains with the moments lying at 120$\degree$ (60$\degree$) from each other. The calculated variation of \pone\ (dashed-dotted line) and \ptwo\ (dotted line) from the magnetic domains cannot reproduce the experimental results. Fitting curves are simulated assuming 67 \% linearly polarized incident light, determined from direct beam measurement. Similar results, corresponding to the $ab$-plane spiral, are obtained when measured at a sample temperature of 14~K. 

\begin{figure}[h!bt]
\centering
\includegraphics[angle=0,width=0.5\textwidth,trim=1cm 1cm 0cm 1cm, clip=true]{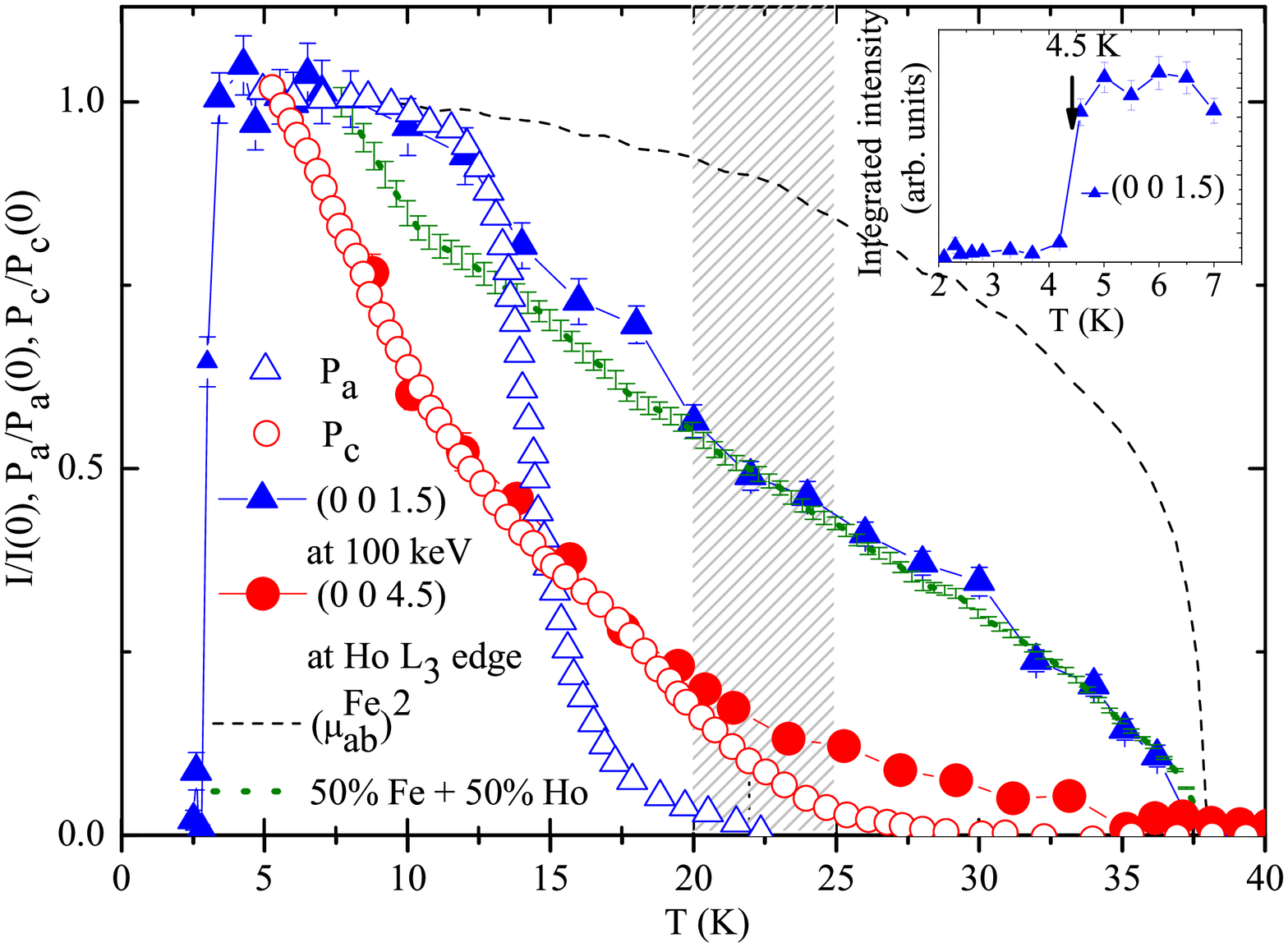}
\caption{(Color online) Temperature dependences of the normalized intensities measured at the
magnetic (0 0 1.5) reflection using 100 keV photon energy (solid triangles) and at the
magnetic (0 0 4.5) reflection at the Ho \lthree\ resonance (solid circles) in \sipi. The
inset shows the region around \tsr, measured with attenuated flux to reduce beam heating. Hollow symbols are the electric polarization data, in the basal plane (triangles) and along the crystal axis (circles), from Ref.~\cite{Cha09}. Dashed and dotted lines show the squared Fe $ab$ plane magnetic moments from Ref. \cite{Rit08} and the expected signal for 50 \% Fe + 50 \% Ho, respectively. The $y$ axis of the data are normalized to the (extrapolated) saturation values.}
\label{tdep}
\end{figure}

The integrated HEXMS intensity of the (0 0 1.5) reflection measured at 100 keV is shown as a function of temperature in Fig.~\ref{tdep} together with the integrated intensities of the (0 0 4.5) reflection measured as a function of temperature in the \sipi\ channel at the Ho resonance. The completely different behavior of the intensities with temperature is obvious. This can be explained by the element specifity of RXS on one hand and by the simplified magnetic scattering cross-section at high energies \cite{Str96} on the other hand. The intensity of the (0 0 4.5) reflection at the Ho \lthree-edge results from the ordering of the Ho moments. The intensity evolution  of the Ho RXS signal with decreasing temperature is characteristic of the polarization of the Ho moments by the ordering of the Fe moments. This indicates that the magnetic scattering in the resonant condition is due to an induced Ho moment. As described in Ref. \cite{Str96}, for small scattering angles HEXMS probes the component of the spin projected onto the normal of the scattering plane i.e. along $\boldsymbol U_{2}$ in Fig.~\ref{scattgeo}. The sample was mounted so that the $ab$ plane, specifically the [h -h 0] direction, is oriented perpendicular to the diffraction plane and hence moments in the $ab$ plane along [h -h 0] are probed. In order to have large enough intensities, measurements were performed at the lowest possible position in $Q$, (0 0 1.5), since the non-resonant X-ray scattering cross-section includes the magnetic form factor, which decreases rapidly with increasing $Q$~\cite{Str96}. At 100 keV photon energies, the scattering angles are small, i.e. 1.41 and 4.23 degrees for the (0 0 1.5) and (0 0 4.5) reflections, respectively. Therefore, an increase in the HEXMS intensity e.g. with temperature of any reflection along [0, 0, l] directly indicates an increasing magnetic moment perpendicular to the scattering plane. For the $ab$-plane Ho spin spirals, the intensities measured at resonance and off-resonance are proportional to the square of the magnetization, so that the normalized intensities can be directly compared. Therefore, the HEXMS intensity of the (0 0 1.5) reflection consists of a combination of Ho and Fe moments, which have large components in the $ab$ plane. The complete vanishing of the HEXMS signal at \tsr=4.5 K, as shown in the inset of Fig.~\ref{tdep}, means that all Fe as well as Ho moments undergo a spin-reorientation, easy plane to easy axis, below \tsr. We would like to mention that in neutron diffraction \cite{Rit08}, below \tsr, Ho moments are observed as oriented by 60$\degree$ with respect to each other along the $c$ axis, instead. The small $ab$-component of the Ho moments below \tsr\ in neutron diffraction measurements has been determined from an additional magnetic peak appearing below \tsr~\cite{Rit08}. We were unable to measure this reflection at the Ho \lthree\ resonance due to beam heating, and it was too weak to be detected by HEXMS. At this point we can conclude that the ordering of the Fe moments in the $ab$ plane is probed through non-resonant magnetic X-ray scattering, and the observation of zero polarization \cite{Cha09} below \tsr\ is due to nearly coinciding moment directions of Fe and Ho. 

In the HEXMS measurement we see a combination of Ho and Fe moments, as both moments lie in the $ab$ plane. To model the two contributions, we sum 50\% of the signal from Ho taken from RXS measurement and 50\% of the signal from the $ab$-plane component of Fe taken from Ref.~\cite{Rit08}, normalized to their saturation values at 5~K. The 50\%-50\% ratio agrees with saturation magnetization values of about 5 $\mu_{B}$ for both magnetic sublattice~\cite{Rit08}. The simulated intensity is in agreement with the experimental data down to 22.5~K (see Fig.~\ref{tdep}). However, to fit the HEXMS data further below 22.5~K, the Fe contribution must be increased significantly. This clearly indicates that at about 22.5 K the Fe moments rotate within the $ab$ plane out of the scattering plane. The total moment in the basal plane remains unchanged for a rotation of the moments within the basal plane. Neutron powder diffraction can not observe a rotation of the Fe spins in the basal plane in this compound, since neutron diffraction is sensitive to the total magnetic moment perpendicular to $Q$ i.e. moments in the basal plane in the present case. This is different for HEXMS, where the moment component perpendicular to the diffraction plane contributes to the magnetic signal. An Fe spin rotation within the $ab$-plane seems to be directly connected with the second step of the two step variation in the Ho RXS signal, (i) a slow increase down to $\sim$ 23 K and then (ii) a fast up-rise below this temperature. Ritter et al.~\cite{Rit08} also observe a rapid increase of the Ho $ab$ plane magnetic moment and a slow increase of the $c$ component of the Fe moments below $\sim$ 23 K. 

The magnetic scattering intensity is proportional to the magnetization squared, $I= AM^{2}$, and in the case of spiral magnets spontaneous electric polarization $\boldsymbol P$~$\propto M^{2}$ \cite{Mos06,Hos05}. These can thus directly be compared ($I$ and $\boldsymbol P$). The temperature dependences of HEXMS and RXS intensities in the basal plane are very similar to the spontaneous electric polarization measured by Chaudhury et al.~\cite{Cha09}. Zero field electric polarization data in warming cycles (normalized to the saturation values) along the $a$ axis and the $c$ axis from Ref.~\cite{Cha09} are plotted together with our measurements (see Fig.~\ref{tdep}). The close relationship strongly suggests that the spontaneous electric polarizations arising along the $a$ and $c$ axes, are due to distinct contributions coming from the Fe and Ho magnetic ordering, respectively. 

It should be noted that the inverse Dzyaloshinskii-Moriya (DM) interaction ~\cite{Dzi58,Mor60} which has been used to explain the direction of $\boldsymbol P$ in the spiral magnets would not be valid in this case. According to the inverse DM interaction, $\boldsymbol P$~$\propto$~e$_{ij}$~$\times$~(S$_{i}$~$\times$~S$_{j}$), where S$_{i}$ and S$_{j}$ are the spin moments on two neighboring magnetic sites and e$_{ij}$ is the unit vector connecting the two sites. For the screw-type magnetic structure e$_{ij}$ being parallel to S$_{i}$~$\times$~S$_{j}$ should result into $\boldsymbol P$=~0. Numerous other examples of systems exist in which the inverse DM interaction does not drive the ferroelectric polarization. For example in HoMnO$_{3}$, the origin of $\boldsymbol P$ was explained by exchange striction of Ho-Mn (for $\boldsymbol P$ along the $c$ axis) and Mn-Mn (for $\boldsymbol P$ along the $a$ axis) \cite{Lee11}. Meanwhile, very recently it was found that in the helical spin spiral system CaMn$_{7}$O$_{12}$ \cite{Joh12}, $\boldsymbol P$ develops as a result of an axial lattice distortion constituting an axial vector which remains invariant under inversion. Returning to \hofeb, one possible mechanism to explain the observed $\boldsymbol P$, could be a symmetric exchange interaction, similar to that used to explain the anomalously large $\boldsymbol P$ in DyMnO$_{3}$~\cite{Fey10}. According to this mechanism, $\boldsymbol P$ in \hofeb\ would arise through a symmetric exchange interaction between the Ho and Fe moments ordering with the same periodicities, $\tau^{Ho}= \tau^{Fe}= \frac{3}{2}$. This would cause a polar lattice modulation with $q= 0$, inducing spontaneous electric polarization. Also, in the present study, due to a subtle observation of the Fe spin rotation within the $ab$-plane at~$\sim$~23~K, which has been identified as coinciding with the onset of $\boldsymbol P$, \hofeb\ indicate similarities to the $R$MnO$_{3}$ compounds~\cite{Ali08}. In $R$MnO$_{3}$ ($R$= Tb and Dy)~\cite{Ali08}, there exist three distinct phase transitions which are characterized as $1)$~\tn~$\sim$~41~K, where Mn spins order with a propagation vector along the $b$-direction; $2)$~$T_S$~$\sim$~28~K for $R$= Tb and $\sim$~18~K for $R$= Dy, identified as an onset of $\boldsymbol P$, where the Mn spins develop a component along the $c$-axis; and $3)$~a final transition \tn~$\sim$~7~K, where the rare earths order with the wave vector along the $b$ direction. Similar such transitions in \hofeb\ could be identified as $1)$~\tn~$\sim$~39~K, where Ho and Fe both order antiferromagnetically; $2)$~$T_S$~$\sim$~23~K, the onset of $\boldsymbol P$, where Fe spins are found to rotate within the $ab$-plane; and $3)$ a final spin reorientation transition at \tsr~$\sim$~5~K occurring due to the strong easy axis magnetic anisotropy of the Ho sublattice. Though the symmetric exchange interaction mechanism in general seems to respect the observation of polarization in this system, without detailed information of the magnetic structure of the Fe magnetic sublattice it will be highly speculative to attribute the directionalities of the $\boldsymbol P$ to the interactions among specific ions (Ho-Ho, Fe-Fe and Ho-Fe). Nevertheless, we would like to emphasize our observation of a scaling of the Ho squared magnetization with $\boldsymbol P_{c}$ and the Fe squared magnetization with $\boldsymbol P_{a}$. In addition to similarities with the extensively studied $R$MnO$_{3}$ compounds, the direct observation of the screw-type magnetic structure of the Ho sublattice and scaling of the ferroelectric polarization with the separated contributions of the Ho and Fe magnetizations in \hofeb\ open up a new avenue for understanding the multiferroicity mechanism in similar compounds. 

\section{Summary}
In conclusion, Ho moments form an $ab$-plane spin spiral propagating along the $c$ axis i.e.~a screw-type magnetic structure. An accelerated increase of the magnetization of the Ho magnetic sublattice below 22.5~K seems to be a driving force behind the observed rotation of the Fe spins within the basal plane. This temperature is identified as the onset of the spontaneous electric polarization in the system. The Ho RXS signal directly scales with the polarization developed along the $c$ axis \cite{Cha09} below 18 K. However, polarization developed along the $a$ axis \cite{Cha09}, below 15 K, scales with the HEXMS intensity which has been identified as a contribution mostly from Fe magnetic moments. These observations demonstrate that the macroscopic electric polarization in these systems develops in close connection with spin rearrangement within the magnetic ordered phase, due to competition between the different magnetic order parameters and magnetic anisotropies. This study is useful to understand the spontaneous electric polarization in a broad range of multiferroic materials, where, below the onset of $\tn$, magnetization from more than one magnetic sublattice and strong structural and magnetic anisotropies are involved.     

\acknowledgements
The authors would like to thank D. Reuther, R. D\"{o}ring, K. Pflaum and R. Nowak for the help in preparing the experiments.
\smallskip

\bibliography{bib_hofeb}

\end{document}